\documentclass[prl,authorgroup,twocolumn,showpacs,preprintnumbers,amsmath,amssymb]{revtex4}

\usepackage{graphicx}
\usepackage{dcolumn}
\usepackage{bm}

\begin{document}

\preprint{APS/123-QED}

\title{Light dressed-excitons in an incoherent-electron sea: Evidence for
Mollow-triplet and Autler-Townes doublet}

\author{Jean~Berney}
 \affiliation{Institut de Photonique et Electronique Quantiques,
 Ecole Polytechnique F\'{e}d\'{e}rale de Lausanne (EPFL) CH1015
 Lausanne, Switzerland}
\author{Marcia T.~Portella-Oberli}
 \affiliation{Institut de Photonique et Electronique Quantiques,
 Ecole Polytechnique F\'{e}d\'{e}rale de Lausanne (EPFL) CH1015
 Lausanne, Switzerland}
\author{Beno{\^i}t~Deveaud}
 \affiliation{Institut de Photonique et Electronique Quantiques,
 Ecole Polytechnique F\'{e}d\'{e}rale de Lausanne (EPFL) CH1015
 Lausanne, Switzerland}
\date{\today}

\begin{abstract}
We demonstrate that the interaction between excitons and a sea of
incoherent electrons does not preclude excitons dressing by light.
We investigate the role of exciton-electron scattering in the light
dressing by measuring the dynamical absorption spectrum of a
modulation-doped CdTe quantum well, which shows a clear evidence for
significant electron scattering of the excitonic states. We show the
occurrence of dressed and correlated excitons by detecting quantum
coherent interferences through excitonic Autler-Townes doublet and
ac Stark splitting, which evolves to Mollow triplet with gain. We
also evidence the partial inhibition of the electron-exciton
scattering by exciton-light coupling.
\end{abstract}


\maketitle

The absorption spectrum of semiconductor quantum wells is dominated
by the exciton line. The role of many-body interactions and exciton
correlations in the exciton dynamical absorption spectrum has
attracted much interest for several decades. The response of the
exciton absorption to an \emph{incoherent} and thermalized
electron-hole population obtained through non resonant pumping in
the continuum corresponds mainly to a bleaching of the exciton
resonance. This bleaching, which increases with electron-hole
concentration, is dominated by collision broadening
(excitation-induced dephasing) rather than by the saturation of the
oscillator strength due to phase space filling \cite{janke:1996}. In
the regime of \emph{coherent} pumping at the exciton resonance, the
light-field dresses the electron-hole gas and dynamical Stark
splitting has been clearly observed in GaAs quantum wells embedded
\cite{quochi:1998} or not \cite{saba:2000,schulzgen:1999} in a
microcavity. Actually, in this \emph{coherent regime}, the
excitation-induced dephasing is found to be reduced as a result of
the significant decrease of the Coulomb collision rates
\cite{ciuti:2000}. As a consequence, the light-induced exciton
dressing renders this many-body system comparable to a
non-interacting atom system \cite{shore:1990}. This issue is very
important if one tries to extend coherent manipulations as those
performed in atomic systems \cite{fleischhauer:2005} to
semiconductors in view of possible applications.

In modulation-doped semiconductor quantum wells, excess carriers
coalesce with excitons to form charged excitons (trions). The
absorption spectrum is then modified: a trion resonance appears
below the exciton resonance \cite{keng:1993}. However, many-body
interactions also tend to become more complex. Electrons, for
instance, are known to strongly screen the exciton oscillator
strength \cite{keng:1993,huard:2000} and to modify their linewidth
and binding energy through exchange interactions \cite{ramon:2003}.
Thus, the shape of the exciton line, as well as that of the trion
line, depends strongly on the strength of electron scattering with
excitons and trions. Indeed, we have observed, as others before us,
that a high energy tail develops at both exciton and trion
resonances upon increasing electron concentration
\cite{berney:2007}. This gives strong evidence for the strength of
the interaction of electrons with excitons and trions. In this
condition, it is expected that coherent exciton-light coupling
should be destroyed due to electron-exciton collision broadening,
and profoundly change the importance of light dressing on excitons.
Actually, we evidence that the nonlinear exciton-light coupling
dominates the electron-exciton scattering. This fact brings new
openings towards possible coherent manipulations in semiconductor
quantum wells containing an electron gas.

\begin{figure}[b]
    \includegraphics[scale=0.85]{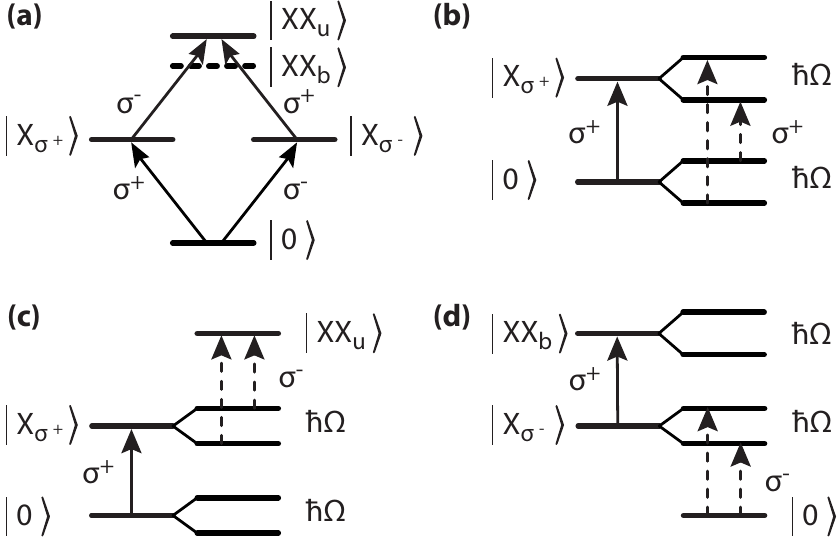}
    \caption{\label{fig:levels}
        Exciton-light coupling diagram, respecting exciton-transition selection rules: (a)  $\sigma^+$ ($\sigma^-$) spin-up +1 (spin-down -1) exciton state $\textrm{X}_{\sigma^+}$ ($\textrm{X}_{\sigma^-}$) with -1/2 (+1/2) electron spin and hole angular momentum projection +3/2 (-3/2). (b) $\sigma^+$ pump pulse dresses $\textrm{X}_{\sigma^+}$ state which is probed by a $\sigma^+$ probe pulse, when it is probed by a $\sigma^-$ pulse, the transition is to unbound $\textrm{XX}_u$ state (c). (d) The bound $\textrm{XX}_b$ is dressed by a $\sigma^+$ pump pulse and the $\textrm{X}_{\sigma^-}$ state is probed by a $\sigma^-$ pulse. $\hbar\Omega$ represents the ac Rabi splitting of the dressed eigenstates.
    }
\end{figure}

In this paper, we investigate the role of light dressing-exciton in
the scattering of excitons with thermalized electrons. We selected a
sample which evidences electron scatterings with excitons and trions
through a prominent high energy tail in their absorption resonances.
We use a pump pulse to coherently prepare the exciton state and we
follow the evolution of the excitonic absorption line by measuring
the spectrum of a weak probe pulse at different delays with respect
to the preparing pump pulse. Detecting the excitonic ac Stark
splitting, Mollow triplet and Autler-Townes doublet allows us to
demonstrate that the interaction between thermalized electrons and
excitons does not prevent dressing excitons. In addition, we
evidence that exciton-light coupling partially inhibits scattering
of excitons by electrons. This is established by exciton resonance
narrowing and reduction of the high energy absorption tail.

Possible exciton transitions are determined by the optical selection
rules. In Fig.~1(a), we show the diagram for exciton-light coupling
for pulsed  $\sigma^+$ and  $\sigma^-$ circular-polarizations,
considering both excitons (X) and biexcitons bound ($\textrm{XX}_b$)
and unbound ($\textrm{XX}_u$) states. The heavy-hole exciton state
$\textrm{X}_{\sigma^+}$ ($\textrm{X}_{\sigma^-}$) is dressed by a
$\sigma^+$ ($\sigma^-$) pump pulse and probed by either a $\sigma^+$
(Fig.~1(b)) or a $\sigma^-$ (Fig.~1(c)) probe pulse. Fig.~1(d) shows
the scheme for dressing the bound-biexciton state $\textrm{XX}_b$
and probing the exciton state $X_{\sigma^-}$.

We worked on a one-side modulation doped
$\textrm{CdTe/Cd}_{0.27}\textrm{Mg}_{0.73}\textrm{Te}$
heterostructure \cite{wojtowicz:1998}, containing a single CdTe
quantum well of 8~nm yielding absorption-like reflectivity spectra
\cite{berney:2007}. A remote donor layer of iodine was embedded in
the cap layer, yielding a population of about
$4\times10^{10}$~cm$^{-2}$ excess electrons in the QW. In this
regime, at 5~K, two resonances separated by 3~meV arise below the
band gap; they are attributed to heavy-hole excitons (1625.7~meV)
and negatively charged excitons (trions, 1622.4~meV).

At the bottom of Fig.~2(a) and 2(b) we plot the reflectivity spectra
of the sample at -20~ps delay time, which is identical to its cw
reflectivity spectrum \cite{berney:2007}. It is important to notice
the high energy tail of both exciton and trion resonances. We
performed a calculation in which we have properly included
electron-trion and electron-exciton scattering. It allows us to
reproduce adequately such high energy tails [see \cite{berney:2007}
for details]. Our calculations further indicate that exchange
scattering is very strong compared to direct Coulomb scattering, as
also shown by others studies \cite{ramon:2003,ciuti:1998a}, and
therefore dominates the broadening mechanisms.

\begin{figure}
    \includegraphics[scale=0.95]{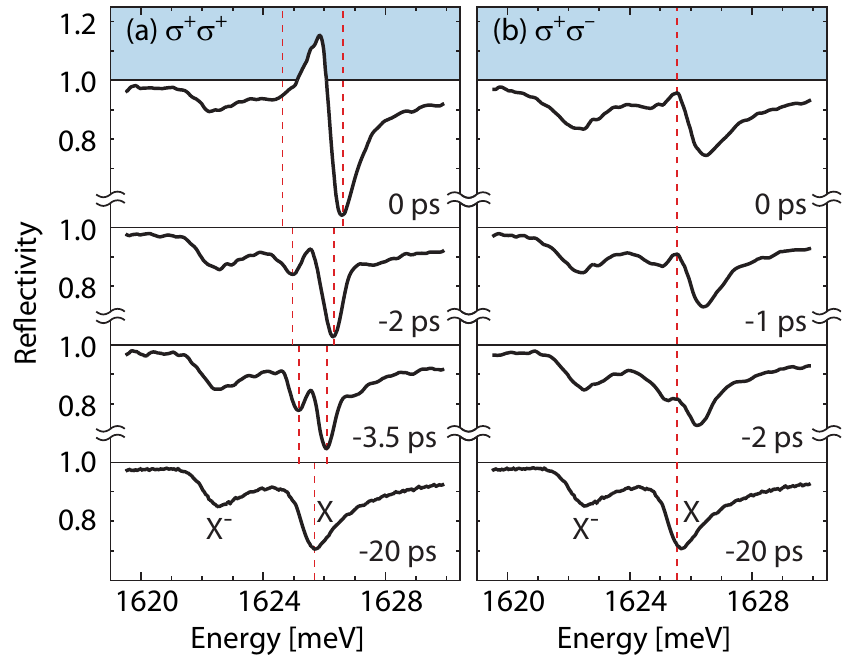}
    \caption{\label{fig:XonR} (color online).
        Reflectivity spectra collected at various delay times for a
        $\sigma^+$ pump tuned to the exciton resonance (1625.7~meV). The vertical scale is the same for all spectra. The probe is co-polarized and the dashed lines indicate the position of the Stark split modes (a). The probe is counter-polarized and the dashed-line indicates the position of the pump pulse
        (b).
    }
\end{figure}

Pump and probe experiments were performed in reflectivity at 5~K.
The 100~fs output pulses of a Ti:Saphire laser were split into two:
a small portion circulated through a delay line and probed the
changes in susceptibility near the exciton resonance, while the
major portion passed through a pulse shaper to generate 4~ps long
and 0.5~meV wide tunable pump pulses. The pump/probe intensity ratio
was larger than 50 for all experiments so that probe reflectivity
spectra remained linear in the probe field. The polarization of both
pump and probe pulses were chosen independently. We recorded the
reflectivity (Fig.~2) and differential reflectivity (Fig.~3 and~4)
spectra for various delays between pump and probe pulses. As usual,
negative delay means that the probe leads the pump.

In our investigations, we use the pump pulse to dress exciton states
and we are interested to follow their evolution in time. It is
important to know that dressed states are best evidenced when the
probe and pump pulses overlap at negative time delays
\cite{quochi:1998, saba:2000}. Then, we use pump pulses with 4~ps
duration and with the highest possible intensity (about
$2\times10^{11}$~photons/pulse cm$^2$).

First, we tuned the pump pulse on the exciton resonance and we
worked in  $\sigma^+\sigma^+$ polarization configuration
(Fig.~1(b)). As the probe delay is changed at negative delays, we
monitored the build up of the pump field and its effect on the
exciton resonance. As Fig.~3(a) shows, the exciton resonance splits
up as the pump intensity rises. Two side modes grow apart, with an
energy separation scaling with the square root of the pump
intensity. This is a key feature of the ac Stark splitting effect,
which originates from a nonlinear coupling between the strong
electromagnetic field and the excitonic resonance. This coherent
light-exciton coupling is characterized by a Rabi energy
$\hbar\Omega$. The Rabi energy is half of the separation between the
sidebands, which correspond to transitions between the eigenstates
of the exciton system dressed by the electromagnetic field
(Fig.~1(b)). Would the exciton-electron scattering be very
efficient, the exciton dephasing time would be very fast thereby
destroying the coherent exciton-light coupling. In such a condition,
the ac Stark splitting would not be observable. The clear
observation of the dressed-exciton in the optical response of a
quantum well containing free carriers indicates that the nonlinear
exciton-light coupling dominates the electron-exciton scattering.
Upon excitation we clearly observe a narrowing of the Stark
sidebands together with a reduction of the high-energy tail. It is
important to remind that this high-energy absorption is due to
exchange electron-exciton scattering. According to this discussion,
our result evidences that the dressing of excitons by light
dramatically reduces their exchange interaction with the thermalized
electrons.

Interestingly enough, the original exciton spectral position becomes
increasingly transparent as the pump power increases and eventually
the transparency turns into gain as in a strongly driven two-level
system. Mollow predicted this gain without inversion of
population~\cite{mollow:1972}; it can be seen as interferences of
various paths of n-photons transitions between dressed states. Ciuti
et al. \cite{ciuti:2000,ciuti:1998b} have shown that this simple
two-level picture still holds for the much more complex case of
excitons in a quantum well, allowing for the existence of gain. This
gain observed here evidences the occurrence of quantum coherent
interferences via dressed exciton states, even in the presence of
thermalized electrons.

Of course, the pump field is also absorbed, which creates excitons.
At zero delay, the estimated density of the photogenerated
$\sigma^+$ polarized exciton gas is $5\times10^{10}$~cm$^{-2}$.
Excitons repel each other due to Pauli exclusion principle through
exchange of their constituent carriers. The exciton resonance
blueshifts \cite{schmitt-rink:1985}, rendering the exciton Stark
split states asymmetric, as we observe. The build up of the exciton
blue shift is clearer on differential reflectivity spectra
(Fig.~3(a)). Similarly to ac Stark splitting, the role of electrons
on exciton-exciton correlations of same spin through exchange
interaction is not preponderant. Still, their signature is very
clear at positive delay: they coalesce with excitons to form trions
within 10~ps~\cite{portella:2003}.

\begin{figure}
    \includegraphics[scale=0.95]{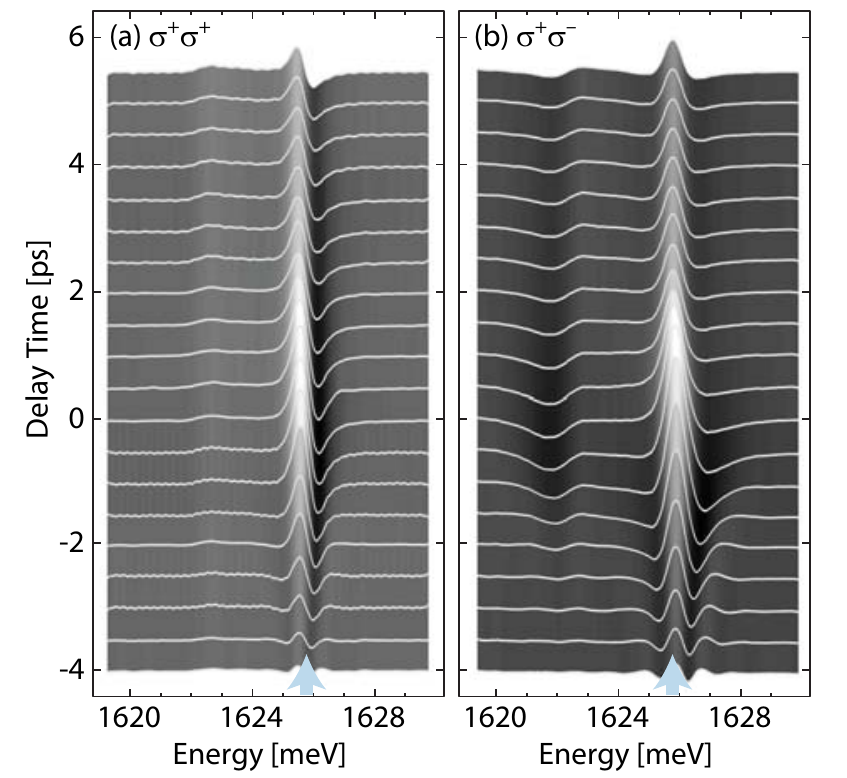}
\caption{ \label{fig:Xon} (color online).
    Differential reflectivity spectra as a function of the
    delay time for (a) $\sigma^+\sigma^+$ and (b) $\sigma^+\sigma^-$ pump and
    probe. The $\sigma^+$ pump (vertical arrow) is tuned to the exciton resonance
(1625.7~meV).}
\end{figure}

Since the  $\sigma^+$ pump field only couples to exciton with
angular momentum $+1$, ac Stark splitting should not be observed for
$\sigma^+\sigma^-$ counter circular polarizations because the
$\sigma^+$ and $\sigma^-$ transitions share no common states
(Fig.~1(a)). However, if excitons with opposite spins are
correlated, they can form bound and unbound biexciton states (see
three-level system in Fig.~1(a)). Then, although  $\sigma^+$ and
$\sigma^-$ transitions share no common state, the  $\sigma^+$ pump
can dress the $\textrm{X}_{\sigma^+}$ state and the probe may
coherently couple this $\textrm{X}_{\sigma^+}$ dressed state to the
unbound biexciton state $\textrm{XX}_u$ (Fig.~1(c)). Indeed, as the
pump field builds up, a spectral dip appears in the probe
reflectivity spectrum at the exciton energy (Fig.~2(b)). At zero
delay, the sample even becomes fully transparent at this wavelength.
So the probe reflectivity spectrum features two resonances called
Autler-Townes doublet \cite{autler:1955}. This doublet may be traced
back to both transitions from $\textrm{X}_{\sigma^+}$ dressed state
to undressed unbound biexciton state $\textrm{XX}_u$ (Fig.~1(c));
the Rabi energy being the separation between the resonances, which
is half the energy separation between the sidebands in ac Stark
splitting. Indeed, we find a smaller energy separation between the
two sidebands for counter-circular polarization than for co-circular
polarization. However, the measured energy is not exactly the
expected value, which is not surprising considering that this
three-level system is formed by correlated exciton and biexciton
states together with their renormalization energies. This result
would require a full theoretical analysis, which is not the scope of
this paper.

Due to the fact that Autler-Townes doublet is detected by probing
the transitions from eigenstates of the dressed exciton to
\emph{undressed} biexciton state, it should be much more sensitive
to the presence of electrons than in the case of ac Stark splitting.
Actually, this is revealed by the tiny decrease of the high energy
tail of their sidebands.

More insight on the exciton correlations is gained by looking at the
differential spectra in Fig.~3(b). The observed nonlinearities on
the exciton resonance at short positive delay times are very much
comparable to the ones observed for  $\sigma^+\sigma^+$ pump and
probe Fig.~3(a); a blueshift of the exciton resonance is clearly
visible as well as a strong bleaching of the absorption line. Yet,
neither Pauli blocking nor first-order Coulomb-induced nonlinearity
are expected to lead to a coupling among the subspaces of different
exciton spin state, thus, the observed correlations evidence
high-order Coulomb correlations between excitons \cite{meier:2000}.
Their strength makes the observation of biexciton bound states
likely \cite{neukirch:2000} and we associate the induced absorption
that shows up about 4~meV below the exciton line to bound
biexcitons~\cite{berney:2007}.

Since bound biexciton states can be formed within an electron gas,
we can now determine if, similarly to unbound biexciton states, they
may take part into coherent processes. To this aim, we tune the pump
close to the bound-biexciton resonance in  $\sigma^+\sigma^-$
configuration \cite{note}. At exciton resonance and negative delay
times, differential spectra (Fig.~4(b)) evidence a coherent signal
similar to the one observed in Fig.~3(b). Despite the electron
population, the virtual bound-biexciton state $\textrm{XX}_b$ is
dressed by the $\sigma^+$ pump pulse and the Autler-Townes doublet
can be probed via the $\textrm{X}_{\sigma^-}$ exciton transition
(Fig.~1(d)). This is possible because the $\textrm{X}_{\sigma^-}$
exciton and the $\sigma^+$ dressed $\textrm{XX}_b$ biexciton share a
common state. In the case of $\sigma^+\sigma^+$ configuration, bound
biexcitons cannot be formed and no coherent signal is established at
exciton resonance (Fig.~4(a)).

\begin{figure}
  \includegraphics[scale=0.95]{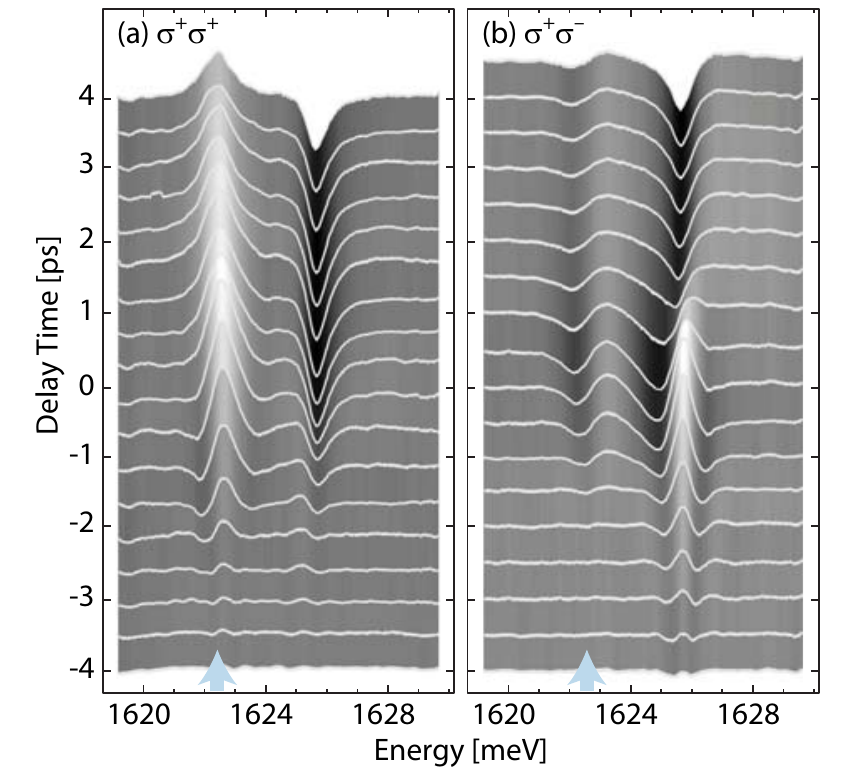}
\caption{\label{fig:Xoff}
    (color online). Differential reflectivity spectra as a function of the
    delay time for (a) $\sigma^+\sigma^+$ and (b) $\sigma^+\sigma^-$ pump and
    probe. The $\sigma^+$ pump (vertical arrow) is tuned close to the biexciton resonance (1621.8~meV).}
\end{figure}

This result is corroborated by the observation, close to zero delay,
of the \emph{excitonic optical Stark shift}
\cite{mysyrowicz:1986,hulin:1990,combescot:1988,combescot:1989},
which is due to a coupling between the exciton and all virtual
two-exciton states, bound and unbound. At small detuning, when the
pump is tuned close to biexciton energy, this coupling is manly due
to Coulomb interaction. It leads to a red-shift of the exciton line
in the case of stable bound biexciton state \cite{combescot:1989},
otherwise the exciton resonance undergoes a blue-shift. Indeed, the
exciton resonance is blue-shifted in $\sigma^+\sigma^+$
configuration (Fig.~4(a)), while in $\sigma^+\sigma^-$ it is
red-shifted (Fig.~4(b)).
%

When the pump is tuned close to the biexciton resonance, it overlaps
inevitably the trion resonance. Thus, a population of
$2\times10^{10}$~cm$^{-2}$ trions is generated, depleting the well
from its excess electrons. It diminishes the scattering of the
excitons by electrons leading to a net induced absorption at the
exciton resonance with a bleaching at the high energy tail
\cite{portella:2004}. This results evidences that the presence of
electrons provides efficient screening of excitons, attenuating
their oscillator strength and broadening their resonance.
Nevertheless, light dressing of excitons, quantum coherence and
exciton correlations are omnipresent and still observed.

As a conclusion, we evidenced that exciton coupling to the incident
field remains sufficiently robust in the presence of electrons to
observe ac Stark splitting, which evolves in a Mollow triplet with
gain at exciton resonance. In addition, we demonstrated that exciton
correlations remain very strong within an electron gas. They yield
unbound and stable bound biexciton pairs that play an important role
in quantum coherent processes evidenced via Autler-Townes doublet
and optical Stark shift. We also showed a narrowing of the exciton
sidebands, evidencing the lowering of scattering between
incoherent-electrons and dressed-excitons.

Acknowledgments: We acknowledge financial support from Swiss
National Science Foundation. We thank M. A. Dupertuis for
enlightening discussions and critical reading of the manuscript. We
are indebted to M. Kutrowski and T. Wojtowicz for growing our
sample.


\begin{thebibliography}{99}
\bibitem{janke:1996} F.~Janke \textit{et al.}, Phys. Rev. Lett. \textbf{77}, 5257 (1996).
\bibitem{quochi:1998} F.~Quochi \textit{et al.}, Phys. Rev. Lett. \textbf{80}, 4733 (1998).
\bibitem{saba:2000} M.~Saba  \textit{et al.}, Phys. Rev. B \textbf{67}, R16322 (2000).
\bibitem{schulzgen:1999} A. Schulzgen \textit{et al.}, Phys. Rev. Lett. \textbf{82}, 2346 (1999).
\bibitem{ciuti:2000} C. Ciuti \textit{et al.}, Phys. Rev. Lett. \textbf{84}, 1752 (2000).
\bibitem{shore:1990} B.~W.~Shore, The Theory of Coherent Atomic Excitation, Vol. 1, Wiley, New York (1990).
\bibitem{fleischhauer:2005} M.~Fleischhauer~\textit{et al.}, Rev. Mod. Phys. \textbf{77}, 633 (2005).
\bibitem{keng:1993} K.~Keng \textit{et al.}, Phys. Rev. Lett. \textbf{71}, 1752 (1993).
\bibitem{huard:2000}  V.~Huard \textit{et al.}, Phys. Rev. B \textbf{84}, 187 (2000).
\bibitem{ramon:2003} G.~Ramon \textit{et al.}, Phys. Rev. B \textbf{67}, 45323 (2003).
\bibitem{berney:2007} J.~Berney \textit{et al.}, Exciton Correlations within an Electron Gas, EPFL, Lausanne (2007).
\bibitem{wojtowicz:1998} W.~Wojtowicz \textit{et al.}, Appl. Phys. Lett. \textbf{73}, 1379 (1998).
\bibitem{ciuti:1998a} C.~Ciuti \textit{et al.}, Phys. Rev. B \textbf{58}, 7926 (1998).
\bibitem{mollow:1972} B.~R.~Mollow \textit{et al.}, Phys. Rev. A \textbf{5}, 2217 (1972).
\bibitem{ciuti:1998b} C.~Ciuti \textit{et al.}, Solid State Comm. \textbf{58}, 7926 (1998).
\bibitem{schmitt-rink:1985} S.~Schmitt-Rink \textit{et al.}, Phys. Rev. B. \textbf{32}, 6601 (1985).
\bibitem{portella:2003} M. T.~Portella-Oberli \emph{et. al}, Phys. Stat. Solidi (b) \textbf{238}, 513 (2003).
\bibitem{autler:1955} S.~H.~Autler and C.~H.~Townes, Phys. Rev. \textbf{100}, 703 (1955).
\bibitem{meier:2000} T.~Meier, Phys. Rev. B \textbf{62}, 4218 (2000).
\bibitem{neukirch:2000} U.~Neukirch, Phys. Rev. B \textbf{61}, R7835 (2000).
\bibitem{note} When off resonant, nonlinearities are small and it is more convenient to assess them throught the study of the differential refletivity spectra.
\bibitem{mysyrowicz:1986} A.~Mysyrowicz \textit{et al.}, Phys. Rev. Lett. \textbf{56}, 2748 (1986).
\bibitem{hulin:1990} D.~Hulin and M.~Joffre, Phys. Rev. Lett. \textbf{65}, 3425 (1990).
\bibitem{combescot:1988} M.~Combescot and R.~Combescot, Phys. Rev. Lett. \textbf{61}, 115 (1988).
\bibitem{combescot:1989} M.~Combescot and R.~Combescot, Phys. Rev. B \textbf{40}, 3788 (1989).
\bibitem{portella:2004} M.~T.~Portella-Oberli \textit{et al.}, Phys. Rev. B \textbf{69}, 235311 (2004); Acta Phys. Pol. A \textbf{106}, 423 (2005).
\end{thebibliography}
\end{document}